\documentclass[twocolumn,prb,showpacs]{revtex4}
\usepackage{psfrag}
\usepackage{graphicx}
\usepackage{amsmath}
\usepackage{amssymb}

\begin{document}
\title{Molecular transport calculations with Wannier functions}
\author{K. S. Thygesen}
\affiliation{Center for Atomic-scale Materials Physics, \\
Department of Physics, Technical University of Denmark, DK - 2800 Kgs. Lyngby, Denmark}
\author{K. W. Jacobsen}
\affiliation{Center for Atomic-scale Materials Physics, \\
Department of Physics, Technical University of Denmark, DK - 2800 Kgs. Lyngby, Denmark}

\date{\today}

\begin{abstract}
  We present a scheme for calculating coherent electron transport in
  atomic-scale contacts. The method combines a formally exact Green's
  function formalism with a mean-field description of the electronic
  structure based on the Kohn-Sham scheme of density functional
  theory.  We use an accurate plane-wave electronic structure method
  to calculate the eigenstates which are subsequently transformed into
  a set of localized Wannier functions (WFs). The WFs provide a highly
  efficient basis set which at the same time is well suited for
  analysis due to the chemical information contained in the WFs.  The
  method is applied to a hydrogen molecule in an infinite Pt wire and
  a benzene-dithiol (BDT) molecule between Au(111) surfaces.  We show
  that the transmission function of BDT in a wide energy window around
  the Fermi level can be completely accounted for by only two
  molecular orbitals.
\end{abstract}
\maketitle

\begin{section}{Introduction}
\label{sec:intro}

The transport of electrons in molecules plays a major role in many
scientific areas. In biological systems for example electrons have to be
transported to or from chemically active parts of enzymes. In
electrochemistry the functioning of an electrochemical cell can be
affected by the electron transport through molecular layers at the
electrodes.  In molecular electronics the goal is to control the
electron transport in such detail that electronic components like
rectifiers or transistors can be constructed from single molecules.
Within these different fields a basic understanding of many phenomena
related to electron transport has been obtained, but the challenge
still remains to develop a detailed, quantitative approach to
accurately determine the transport properties from basic quantum
mechanical principles.

In this paper we address this issue in the domain where the
electronic motion can be regarded as coherent.  We present a numerical
Green's function method for calculating the linear response
conductance of molecules and nano-contacts.  The system under
investigation is divided into a central scattering region
and attached leads and the electronic structure of both the scattering
region and the leads is described using Density Functional Theory
(DFT).  To represent the electronic states we use a set of localized
basis functions consisting of partly occupied Wannier functions (WFs).
The WFs describe the relevant states very accurately -- in fact they are
constructed so that they reproduce the results of
high-accuracy plane-wave based pseudopotential calculations -- and at
the same time they represent a minimal basis set which is
computationally very efficient. 

In addition to their technical advantages, the Wannier functions
provide insight into the local chemistry of the system and are thus
well suited for analysis purposes. Being the local analogue of the
extended (Bloch) states of solid state physics, the Wannier functions
formalize chemical concepts such as bond types and electron lone
pairs.  The traditional definition of Wannier functions for single
isolated bands (insulators) or occupied molecular orbitals (molecules)
does not apply to metallic systems~\cite{iannuzzi02}. In this case an
extended scheme involving the inclusion of selected unoccupied states
through a disentangling procedure must be used instead. We present a
newly developed scheme which solves this problem in a simple and very
direct way and we show how to link the resulting "partly occupied WFs"
to the general transport formalism through the evaluation of the
relevant Hamiltonian matrices.

We apply the approach to two different systems. In the first we
calculate the conductance of a hydrogen molecule suspended between two
monatomic Pt wires. This can be viewed as a simple model of a recent
experiment which investigated the conductance of a molecular hydrogen
contact between bulk Pt electrodes~\cite{smit02}.  By diagonalizing
the Hamiltonian within the subspace spanned by the WF of the molecule
we obtain renormalized bonding and anti-bonding $\text{H}_2$ orbitals.
The individual contributions of these orbitals to the total
transmission is then studied by directly removing each of the orbitals
from the basis set.  In this way we find that transport is due to the
anti-bonding state.  The second system consists of a benzene-dithiol
molecule between Au(111) surfaces. This system was among the first
single-molecule systems to be studied experimentally~\cite{reed97} and
represents a test system for transport calculations. Our results for
the conductance agree qualitatively with previous calculations, but
using the analysis technique described above, we furthermore show that
the transport properties of BDT can be completely accounted for by
only two molecular orbitals.

The paper is organized as follows. In Sec.~\ref{sec.formalism} we
introduce the concept of phase-coherence and give an overview of
existing methods for transport calculations. We then proceed to
develop the general Green's function formalism underlying the present
scheme, and illustrate it through a detailed discussion of the simple
case of transport through a single energy level. In Sec.~\ref{sec.WF}
we introduce a numerical method to construct partly occupied Wannier
functions and apply it to an isolated ethylene molecule and an
infinite trans-polyacetylene wire. In Sec.~\ref{sec.hamiltonian} we
connect the Wannier functions to the general transport formalism by
showing how to obtain the relevant Hamiltonian matrices in a Wannier
function basis. We put emphasis on issues related to the combination
of periodic supercells and transport calculations. In
Sec.~\ref{sec.results} we present conductance calculations for a
molecular hydrogen bridge between monatomic Pt wires and a
benzene-dithiol molecule suspended between Au(111) surfaces.
\end{section}

\begin{section}{Transport theory}\label{sec.formalism}
  We begin this section with a brief discussion of the phase-coherent
  transport regime. In order place our method in the large picture we
  give an overview of previously developed transport schemes.  The
  general Green's function formalism and the central conductance
  formula are then introduced in the case of non-interacting electrons
  moving in a static mean-field potential. Extension of the formalism
  to include ultrasoft pseudopotentials is shown to be
  straightforward. Finally, in order to illustrate the general theory
  we give a detailed discussion of electron transport in the simple
  but important Newns Anderson model.

\begin{subsection}{Phase-coherent electron transport}
  Electron transport is said to be phase-coherent if the dimensions of
  the system under study are smaller than the phase relaxation length,
  which is the length over which the phase of an electron is
  randomized due to scattering~\cite{datta_book,imry_book}. The
  phase-relaxation length thus defines the length scale on which
  electron interference can be observed. A typical phase relaxation
  length for Au at $T=1$~K is around
  $1\mu\text{m}$~\cite{agrait_report}. In general, the phase of an
  electron is destroyed by interactions with the dynamic environment
  such as the other electrons, phonons and magnetic impurities and
  consequently the phase relaxation length increases at lower
  temperatures. Static scatterers such as the potential from the ions
  in a rigid lattice do not destroy the phase since the scattering is
  elastic and the phase change is fixed.
\end{subsection}

\begin{subsection}{Existing transport schemes - an
  overview}\label{sec.methods}
During the last decade a variety of numerical methods addressing the
transport of electrons in atomic-scale systems have been developed.
Very generally, these methods can be divided into wave function and
Green's function methods. In both cases the description is based on a
Landauer-B{\"u}ttiker setup where a central scattering region is
placed between two ballistic leads which are assumed to be in thermal
equilibrium far from the central region.

In the wave function method one solves directly for the scattering
wave functions of the system and obtain the conductance/current from the elastic transmission coefficients 
in accordance with the well known Landauer formula~\cite{buttiker85}.  The scattering
states can be calculated using either the transfer matrix
method~\cite{sautet88,emberly98,hirose94,hirose95,choi99,kobayashi00},
the wave function matching method~\cite{khomayakov04} or by solving a
Lippman-Schwinger equation~\cite{lang95,diventra00}.  These techniques
have been combined with various approximations for the electronic
structure of the system, ranging from semi-empirical models like
extended H{\"u}ckel~\cite{sautet88,emberly98} to fully atomistic
descriptions based on the single-particle Kohn-Sham scheme
of density functional theory (DFT)~\cite{choi99,khomayakov04}. Most of
the wave function methods, however, apply an intermediate level of approximation
where only the electronic structure of the central region is resolved in detail
while the leads are modeled by a free electron gas
(jellium)~\cite{hirose94,hirose95,kobayashi00,lang95,diventra00}.

As an alternative to the Landauer-B{\"u}ttiker approach the transport problem
can be formulated with in the non-equilibrium Green's function formalism developed by Keldysh~\cite{keldysh} and
Kadanoff and Baym~\cite{kadanoff_baym}. An  
advantage of the Green's function (GF) formulation is the possibility
of extending the description beyond the single-particle picture by
including e.g. inelastic scattering on vibrations or correlation
effects through self-energies. While
the inclusion of such many-body effects is formally straightforward, the
practical calculation of reliable self-energies from first principles
is a difficult task which only recently has been
addressed~\cite{frederiksen04}. For non-interacting electrons, the GF- and Landauer-B{\"u}ttiker formalisms are 
equivalent~\cite{meirwingreen}. In this case the main effort of the GF method is to
calculate the GF of the central region in the presence of coupling to
the leads. The effect of coupling to the leads is usually incorporated
through non-hermitian self energies~\cite{taylor01,xue01,brandbyge02,thygesen_bollinger03,calzolari04,ke04,heurich02,derosa01,nardelli99,yeyati97,palacios01}.
The various methods mainly differ in the 
representation of the Hamiltonian and self-energy matrices which 
range from tight-binding
models~\cite{nardelli99,yeyati97}, to fully atomistic DFT
descriptions~\cite{taylor01,xue01,brandbyge02,thygesen_bollinger03,calzolari04,ke04}.
Intermediate descriptions treating only
the central region at the atomic level and adopting a
parametrization of the coupling and lead electronic structure have also been
used~\cite{heurich02,derosa01,palacios01}.

The main difference among the fully DFT based GF methods lies in the
basis set used to evaluate the Hamiltonian and thus the Green's
function. One limitation inherent in the GF formalism is that the
basis set should consist of functions with finite support. As a
consequence most methods have adopted a basis set of numerical orbitals such as
Gaussians~\cite{xue01}, LCAO~\cite{taylor01,brandbyge02,ke04} or wavelets~\cite{thygesen_bollinger03}, for which the condition of finite support can be
exactly fulfilled. In contrast, there exists only a few GF transport
schemes based on the more accurate plane-wave DFT methods~\cite{thygesen_bollinger03,calzolari04}. This is
partly due to the delocalized nature which excludes the
plane waves themselves as
basis functions and partly due to the large number of plane waves
needed to perform even a small calculation with a reasonable
accuracy. As we demonstrate in this paper, both of these problems can be
overcome by transforming the set of pre-calculated eigenstates into
an equivalent set of localized Wannier functions. 
\end{subsection}

\begin{subsection}{Conductance formula}
  We consider the phase-coherent transport of electrons through a
  system of the form sketched in Fig.~\ref{fig.systemsetup}. The
  system consists of a scattering region ($S$) connected to thermal
  reservoirs via left ($L$) and right ($R$) metallic leads. The
  reservoirs have a common temperature but different electro-chemical
  potentials $\mu_L$ and $\mu_R$, respectively. We shall assume that
  the electrons are non-interacting and move in a static mean-field
  potential.  The leads are assumed to be perfect conductors and thus
  the electrons move ballistically in these regions and can scatter
  only on the potential inside $S$.  This assumption has important
  computational consequences which we explore in
  Sec.~\ref{sec.coupling}.

\begin{figure}
\includegraphics[width=0.9\linewidth]{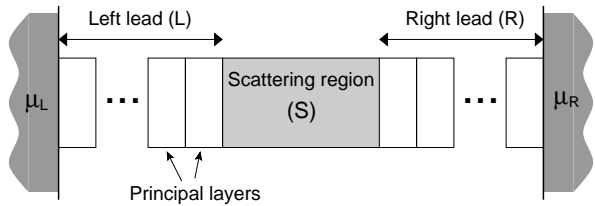}
\caption[cap.Pt6]{\label{fig.systemsetup} Schematic setup used to
  study phase-coherent electron transport. The system is divided into
  three regions: a scattering region ($S$) and two leads ($L$) and ($R$)
  connecting $S$ to thermal reservoirs with chemical
  potentials $\mu_L$ and $\mu_R$, respectively. The electron
  mean-field potential in the leads is periodic and thus all
  scattering takes place in $S$. Each lead can be build from a
  principal layer containing an integer number of potential periods.}
\end{figure}

By introducing a basis set, $\{\phi_i\}$, consisting of functions
with finite support in the direction of transport we can decompose the 
underlying Hilbert space into disjoint subspaces corresponding to the
three regions $S,L,R$. We do not require that the basis functions are orthogonal.
The Hamiltonian matrix, $\boldsymbol{H}_{ij}=\langle \phi_i|H|\phi_j\rangle$, of the entire system takes the form
\begin{equation}
\label{eqn:scatter_hamiltonian}
\boldsymbol{H}=\left(
                \begin{array}{ccc}
                \boldsymbol{H}_{L} & \boldsymbol{H}_{SL}^{\dagger} & 0 \\
                \boldsymbol{H}_{SL} & \boldsymbol{H}_{S} & \boldsymbol{H}_{SR} \\
                0 & \boldsymbol{H}_{SR}^{\dagger} & \boldsymbol{H}_{R} \\
                \end{array}
               \right),
\end{equation}
where the $\boldsymbol{H}_{i}$ themselves are matrices. The vanishing coupling between the leads can always be obtained by
increasing the size of the scattering region, i.e. by including part
of the leads in $S$. It
should be noted that if the basis is not orthonormal then in general
$\boldsymbol{H}$ will be different from the matrix representing the
Hamiltonian in that basis.

For non-interacting electrons the retarded single particle Green's
function can be obtained from the resolvent of the Hamiltonian,
$(zI-H)^{-1}$. We define the Green's function matrix by
\begin{equation}\label{eqn:gf_def1}
(z\boldsymbol{S}-\boldsymbol{H})\boldsymbol{G}^r(\varepsilon)=\boldsymbol{I},
\end{equation}
where $\boldsymbol{S}_{ij}=\langle\phi_i|\phi_j\rangle$ is the overlap matrix, $\boldsymbol{I}$ the
identity matrix and $z=\varepsilon+i\eta$, $\eta$ being a positive
infinitesimal. To emphasize the division of the system into the three
regions we write out Eq.~(\ref{eqn:gf_def1}) more explicitly
\begin{widetext}
\begin{equation}
\label{eqn:gf_def}
\left(
        \begin{array}{ccc}
        z\boldsymbol{S}_{L}-\boldsymbol{H}_{L} 
      & z\boldsymbol{S}_{SL}^{\dagger}-\boldsymbol{H}_{SL}^{\dagger} 
      & 0 \\
        z\boldsymbol{S}_{SL}-\boldsymbol{H}_{SL}
      & z\boldsymbol{S}_{S}-\boldsymbol{H}_{S}
      & z\boldsymbol{S}_{SR}-\boldsymbol{H}_{SR} \\
        0 
      & z\boldsymbol{S}_{SR}^{\dagger}-\boldsymbol{H}_{SR}^{\dagger}
      & z\boldsymbol{S}_{R}-\boldsymbol{H}_{R}
        \end{array}
\right)
\left(
        \begin{array}{ccc}
        \boldsymbol{G}_{L}^r & \boldsymbol{G}_{LS}^r & \boldsymbol{G}_{LR}^r \\ 
        \boldsymbol{G}_{SL}^r & \boldsymbol{G}_{S}^r & \boldsymbol{G}_{SR}^r \\
        \boldsymbol{G}_{RL}^r & \boldsymbol{G}_{RS}^r & \boldsymbol{G}_{R}^r 
        \end{array}
\right)
=
\left(
        \begin{array}{ccc}
        \boldsymbol{I}_{L} & 0 & 0 \\ 
        0 & \boldsymbol{I}_{S} & 0 \\
        0 & 0 & \boldsymbol{I}_{R}  
        \end{array}
\right),
\end{equation}
\end{widetext}

It should be noted that $\boldsymbol{G}^r$ differs from the matrix
 $\langle \phi_i|(zI-H)^{-1}|\phi_j\rangle$ if the
 basis is not orthonormal. The two are related by
\begin{equation}
\label{eqn:gf_relation}
 \langle \phi_i|(zI-H)^{-1}|\phi_j\rangle= \left[\boldsymbol{S}\boldsymbol{G}^r
(\varepsilon)\boldsymbol{S}\right]_{ij}.
\end{equation}
The Green's function of the scattering region, $\boldsymbol{G}_{S}^r$, plays a central role
in the theory. It can be written~\cite{datta_book}
\begin{subequations}
\label{eqn:scattergf}
\begin{eqnarray}
\boldsymbol{G}^{r}_{S}(\varepsilon)&=&\left(
   z\boldsymbol{S}_{S}-\boldsymbol{H}_{S}-
   \boldsymbol{\Sigma}^r_{L}(\varepsilon)-
   \boldsymbol{\Sigma}^r_{R}(\varepsilon)
                                       \right)^{-1}, \label{eqn:scattergf_a} \\
\boldsymbol{\Sigma}^r_{\alpha}(\varepsilon)&=&
   \left(z\boldsymbol{S}_{S\alpha}-
                      \boldsymbol{H}_{S\alpha}\right)
   \boldsymbol{g}^{0,r}_{\alpha}(\varepsilon)(z\boldsymbol{S}_{S\alpha}^{\dagger}-
                    \boldsymbol{H}_{S\alpha}^{\dagger}) 
                       \label{eqn:scattergf_b} \\
\boldsymbol{g}_{\alpha}^{0,r}(\varepsilon)&=&\left(
                      z\boldsymbol{S}_{\alpha}-
                       \boldsymbol{H}_{\alpha}\right)^{-1}, 
                       \label{eqn:scattergf_c}
\end{eqnarray}
\end{subequations}
where $\alpha\in \{L,R\}$. The matrix $\boldsymbol{\Sigma}^r_{\alpha}$
is a non-hermitian self energy which incorporates 
the coupling to lead $\alpha$ and
$\boldsymbol{g}_{\alpha}^{0,r}(\varepsilon)$ is the retarded Green's function
of the uncoupled lead. 
 
An expression for the current through the system was derived by Meir
and Wingreen~\cite{meirwingreen} using
the non-equilibrium Green's function formalism. The
most general version of the formula is also valid when interactions
are present in $S$. However, here we focus on the non-interacting case for which
\begin{equation}
\label{eq.current}
I = \frac{2e}{h}\int \left[n_{\text{F}}(\varepsilon-\mu_{L})-n_{\text{F}}(\varepsilon-\mu_{R})
                           \right]T(\varepsilon)\text{d}\varepsilon,
\end{equation}
where $n_F(\varepsilon)$ is the Fermi distribution function and the
elastic transmission function is given by  
\begin{equation}\label{eq.transmission}
T(\varepsilon)=\text{Tr}\left[
\boldsymbol{G}^{r}_{S}(\varepsilon)
\boldsymbol{\Gamma}_{L}(\varepsilon)
\boldsymbol{G}^{a}_{S}(\varepsilon)
\boldsymbol{\Gamma}_{R}(\varepsilon)
                  \right].
\end{equation}
Here $\boldsymbol{G}^{a}_{S}=[\boldsymbol{G}^{r}_{S}]^{\dagger}$
denotes the advanced Green's function of the scattering region and 
$\boldsymbol{\Gamma}_{\alpha}$ is obtained from the self-energies as
\begin{equation}
\label{eqn:gamma}
\boldsymbol{\Gamma}_{\alpha}(\varepsilon)=i\left(
                    \boldsymbol{\Sigma}_{\alpha}^{r}(\varepsilon)-
                     \left[\boldsymbol{\Sigma}_{\alpha}^{r}(\varepsilon)
                       \right]^{\dagger}\right).
\end{equation}
In the original derivation of Eq.~(\ref{eq.transmission}) the basis set was assumed to be
orthonormal, but in fact the formula holds for a general basis as
has been shown by Xue~\emph{et al.}~\cite{xue01}. Although the 
current formula (\ref{eq.current}) is valid for non-equilibrium
transport we shall focus on the linear response
conductance, $G$. 
Assuming that the difference between $\mu_L$ and $\mu_R$ is
small and taking the zero temperature limit we find 
\begin{equation}
\label{eqn:conductance}
G=G_0 T(\varepsilon_F),
\end{equation}
where the conductance quantum is given by $G_0=2e^2/h$. 
\end{subsection}

\begin{subsection}{Coupling to leads}\label{sec.coupling}
  The expression Eq.~(\ref{eqn:scattergf}) for the self-energy due to lead
  $\alpha$ contains the Green's function
  $\boldsymbol{g}_{\alpha}^{0,r}$ of the uncoupled lead. Since
  transport through the
  leads is assumed to be ballistic the electron potential must be
  periodic in these regions and consequently the leads can be divided
  into principal layers containing an integer number of
  potential periods. Due to the finite range of the basis functions
  in the transport direction, the size of the principal layers can always
  be chosen so large that only neighboring layers couple. In this
  case the Hamiltonian matrix of the left lead takes the form
\begin{equation}
\label{eqn:leadgfeq}
\boldsymbol{H}_L=\left(\begin{array}{cccc}
        \ddots & \vdots & \vdots & \vdots  \\
        \hdots & 
        \boldsymbol{h}_{0} & 
        \boldsymbol{h}_{1} & 0 \\
        \hdots & \boldsymbol{h}_{1}^{\dagger} & 
        \boldsymbol{h}_{0} & 
        \boldsymbol{h}_{1} \\
        \hdots & 0 & \boldsymbol{h}_{1}^{\dagger}& 
        \boldsymbol{h}_{0} \\
        \end{array}
\right)
\end{equation}
where $\boldsymbol{h}_{0}$ is the Hamiltonian matrix of a single
principal layer and $\boldsymbol{h}_{1}$ is the coupling between
neighboring layers. A similar matrix structure is obtained for
$\boldsymbol{H}_R$.  In writing Eq.~(\ref{eqn:leadgfeq}) we have
assumed the lead potential to be periodic all the way up to the
scattering region. This condition can always be fulfilled by extending
the scattering region until it comprises all perturbations arising
from the presence of scatterers. In practice such perturbations are
rapidly screened by the mobile electrons in he metallic leads and the mean field potential
is expected to decay to its bulk value over a few atomic layers.

The division of the leads into principal layers with nearest neighbor
coupling also implies that the scattering region only couples to the
principal layers immediately next to it. From
Eq.~(\ref{eqn:scattergf}) it then follows that only the lead Green's
function of the first principal layer is needed for evaluating the
self-energy. This Green's function can be determined iteratively due
to the periodic structure of the matrix
$(z\boldsymbol{S}_{\alpha}-\boldsymbol{H}_{\alpha})$. A particularly
efficient iteration scheme is provided by the so-called decimation technique~\cite{guinea}. 
\end{subsection}

\begin{subsection}{Ultrasoft pseudopotentials}
So far we have made no assumptions regarding the form of the
mean-field electron potential. In particular it could contain
non-local terms which are a natural component of most
norm-conserving pseudopotentials~\cite{bachelet82,hamann79}. 
However, special care must be taken when ultrasoft
pseudopotentials~\cite{vanderbilt90} are used, as explained below.

Suppose $H$ is an effective single-particle Hamiltonian in which the ion
potential has been replaced by an ultrasoft pseudopotential.
The energy spectrum is found by solving the generalized eigenvalue
equation
\begin{equation}\label{eq.ultrasofteigenvalue}
H\psi_n=\varepsilon_nS\psi_n,
\end{equation}
where the pseudo wave functions satisfy the generalized orthonormality
condition
\begin{equation}
\langle \psi_n|S|\psi_m\rangle=\delta_{nm}.
\end{equation}
Although the Hamiltonian is hermitian its
eigenvalues do not coincide with the real energy spectrum and it is
therefore not obvious at the outset how to define the Green's function
entering the transmission formula (\ref{eq.transmission}).

The operator $\widetilde H=S^{-1}H$ has the correct spectrum but it is not
hermitian - its eigenvectors are not orthogonal. This is, however, not
a real problem since self-adjointness depends on the inner product.
In fact $\widetilde H$ is hermitian if we use the inner product $\langle
\cdot|\cdot\rangle_1$ defined by 
$\langle \psi|\phi\rangle_1=\langle \psi|S|\phi\rangle$. That $\langle
\cdot|\cdot\rangle_1$ does define an inner product
follows from the fact that $S$ is hermitian and positive in the sense $\langle
\phi|S|\phi\rangle\geq 0$ for all $\phi$. 

The apparent problem related to the use of ultrasoft pseudopotentials
can thus be removed by changing the inner product from $\langle
\cdot|\cdot\rangle$ to $\langle
\cdot|\cdot\rangle_1$ and using the Hamiltonian $\widetilde H$ instead
of $H$. A direct application of Eq.~(\ref{eqn:gf_def1}) leads to the
following defining equation for $\boldsymbol{G}^r$:
\begin{equation}
(z\boldsymbol{S}-\boldsymbol{H})^{-1}\boldsymbol{G}^r(\varepsilon)=\boldsymbol{I},
\end{equation}
where $\boldsymbol{S}_{ij}=\langle \phi_i|S|\phi_j\rangle$ and
$\boldsymbol{H}_{ij}=\langle \phi_i|H|\phi_j\rangle$. Consequently, the
only change in the formalism introduced by the use of ultrasoft
pseudopotentials is the substitution of the ordinary overlap matrix
$\langle \phi_i|\phi_j\rangle$ by the generalized overlap $\langle \phi_i|S|\phi_j\rangle$.
\end{subsection}

\begin{subsection}{Newns Anderson model}\label{sec.newns}
In order to illustrate the use of the general formalism introduced above and
to develop a physical understanding of the
conductance formula (\ref{eq.transmission}), we discuss  
the case of transport through a single energy level
coupled to continuous bands. 
As we shall see later this model  
is also useful for analysis and
interpretation of the transport properties of more complicated systems.    

In the Newns Anderson model~\cite{newns69} we consider the single site $|a\rangle$ of
energy $\varepsilon_a$ coupled to infinite leads via the matrix
elements $t_{\alpha \nu}=\langle a|H|\alpha \nu\rangle$, where
$\{|\alpha \nu\rangle\}$ is a basis of lead $\alpha$. The single level
constitutes the scattering region and thus all matrix quantities
in the transmission function~(\ref{eq.transmission}) reduce to complex numbers.
The Green's function reads
\begin{equation}
G_a(\varepsilon)=\frac{1}{\varepsilon-\varepsilon_a-\Sigma_L(\varepsilon)-\Sigma_R(\varepsilon)},
\end{equation}     
with the self-energy due to lead $\alpha$ given by 
\begin{equation}
\Sigma_{\alpha}(\varepsilon)=\sum_{\nu
  \mu}t_{\alpha \mu}(g^0_{\alpha})_{\mu \nu}t_{\alpha \nu}^*,
\end{equation}
We have dropped the $r$ superscripts since we
consider only retarded Green's functions in the following. A particularly elegant solution
to the problem can be obtained by
introducing the normalized group orbital of lead $\alpha$:
\begin{equation}
|\gamma_{\alpha}\rangle=V_{\alpha}^{-1}\sum_{\nu}t_{\alpha \nu}|\alpha \nu\rangle,
\end{equation}
where $V_{\alpha}=(\sum_{\nu} |t_{\alpha \nu}|^2)^{1/2}$. The group
orbital is located on lead $\alpha$ and its weight on a given lead
state is determined by the magnitude of the corresponding coupling
matrix element. Consequently, the group orbital is expected to be
localized at the lead-level interface. The coupling between
$|a\rangle$ and any state in the lead orthogonal to
$|\gamma_{\alpha}\rangle$ vanishes and thus $|a\rangle$ is coupled to
the lead via the group orbital only, the coupling being given by
$V_{\alpha}$. For simplicity we assume a symmetric contact and
therefore suppress the $\alpha$ index. Using the general relation
between a diagonal element of the retarded Green's function and the
projected density of states (DOS) for the corresponding
orbital:~$\text{Im}[G_{\nu \nu}]=-\pi \rho_{\nu}$, we can express
$\Gamma$ (as defined in Eq.~(\ref{eqn:gamma})) by the DOS for the
isolated lead projected onto the group
orbital:~$\Gamma=\pi|V|^2\rho^0_{\gamma}$. Applying the same rule to
the Green's function of the level we arrive at
\begin{equation}\label{eq.newns} 
T(\varepsilon)=2\pi^2|V|^2\rho_a(\varepsilon)\rho^0_{\gamma}(\varepsilon).
\end{equation}
This formula shows that the transmission at a given energy depends on
three quantities: the coupling strength, $V$, the level DOS, $\rho_a$,
and the DOS of the group orbital in the isolated lead,
$\rho_{\gamma}^0$. Thus, for an electron with energy
$\varepsilon$ to propagate through the system there should be states
of that energy available in the leads as well on the level and the
coupling should be reasonably strong. 

An even simpler description is obtained in the so called wide
band limit where $\rho_{\gamma}^0$ is assumed to be constant. In this case
the transmission is usually expressed in terms of $\Gamma$ and
$\varepsilon_a$ and reads
\begin{equation}\label{wideband}
T(\varepsilon)=\frac{\Gamma^2}{(\varepsilon-\varepsilon_a)^2+\Gamma^2}.
\end{equation} 
This is a Lorentzian of unit height and center $\varepsilon_a$. Its
width is given by $\Gamma$ which also determines the width of the energy
distribution of $|a\rangle$ in the coupled system and thus equals the inverse lifetime of
an electron on the level. From this expression it is clear that the position of the Fermi
level relative to $\varepsilon_a$ is a crucial parameter for the
conductance. Finally, we note that deviations from the simple
Lorentzian form (\ref{wideband}) are expected when the coupling is asymmetric or when 
$\rho_{\gamma}^0(\varepsilon)$ has a significant energy dependence.
\end{subsection}
\end{section}

\begin{section}{Wannier Functions}\label{sec.WF}
In this section we present a scheme to construct partly occupied
Wannier Functions (WFs)\cite{thygesen_wannier}. After a brief motivation we outline the
construction algorithm for both isolated and periodic systems. It is
natural to introduce the algorithm first in the simpler case of an isolated system before extending it to periodic
systems, although the latter contains the former as a special case.
For later use we establish an expression for the matrix elements of the
Hamiltonian in the WF basis. Finally, we illustrate the method by
constructing the WFs for an isolated ethylene molecule and an infinite
trans-polyacetylene wire.

\begin{subsection}{Why Wannier functions?}\label{sec.why}
There are several advantages of using WFs as a basis set for
transport calculations\cite{calzolari04}: (i) the WFs are spatially localized which 
allows for the necessary division into scattering region and leads. (ii) any
eigenstate below a certain specified energy can, by construction, be
exactly reproduced as a linear combination of the WFs and thus the
accuracy of the original electronic structure calculation is retained. (iii) the WF
basis set is truly minimal and once constructed the computational cost
of the subsequent transport calculation is comparable to that of
semi-empirical methods like tight-binding or extended H{\"u}ckel. (iv) the WFs contain information about
 chemical properties such as
bond types, coordination and electron lone pairs and can thus be
directly used as an analysis tool.

The WFs are defined as linear
combinations of a fixed set of single-particle eigenstates -
traditionally the occupied states - with the expansion
coefficients optimized to give the best localization of the
resulting WFs. In this paper we shall focus on partly occupied WFs. 
The term "partly occupied" refers to the fact that we
allow selected unoccupied orbitals in the
expansion of the WFs. In some situations this can improve
the symmetry of the resulting orbitals, however, the crucial property
that distinguish the partly occupied WFs from the traditional occupied
WFs and make them applicable for transport calculations is that they can
be localized in metallic systems. It is well known that occupied WFs associated with a partly filled band in a
metal have poor localization properties~\cite{souza01, iannuzzi02}. The
only remedy for this is to include the appropriate unoccupied states in the localization space.
\end{subsection}

\begin{subsection}{Isolated systems}
  Consider an isolated system, for which an electronic structure
  calculation produces $N$ eigenstates,
  $\{\psi_n\}$, of which $M$ have an energy below $E_0$.  Our aim is
  to construct a set of localized orbitals out of the $\{\psi_n\}$, which
  span at least the $M$ eigenstates below $E_0$. We will allow
  eigenstates with energy above $E_0$ in the expansion of the WFs in
  order to improve the localization.

Before we can start to study localized orbitals, we need a 
measure for the degree of localization. Here we follow  
Marzari and Vanderbilt~\cite{marzari97} and measure the
spread of a set of functions, 
$\{w_n\}$, by the sum of second moments:
\begin{equation}\label{eq.spreadfct1}
\mathcal{S}=\sum_{n=1}^N (\langle w_n | r^2 | w_n \rangle - \langle w_n |\mathbf r
| w_n \rangle ^2).
\end{equation}
We expand the WFs in terms of the $M$ lowest eigenstates and $L$
extra degrees of freedom, $\{\phi_l\}$, from the remaining
$(N-M)$-dimensional eigen-subspace. This will lead to
a total of $M+L$ WFs.
The expansion takes the form
\begin{equation}\label{eq.expansion1}
w_n=\sum_{m=1}^{M}U_{mn}\psi_m+\sum_{l=1}^{L}U_{M+l,n}\phi_l,
\end{equation}
where the extra degrees of freedom (EDF) are written as
\begin{equation}\label{eq.expansion2}
\phi_l=\sum_{m=1}^{N-M}c_{ml}\psi_{M+m}.
\end{equation}
The columns of the matrix ${c}$ are taken to be orthonormal and
represent the coordinates of the EDF with respect to the 
eigenstates with energy above $E_0$. The matrix $U$ is unitary and represents a rotation in
the space of the functions
$\{\psi_1,\ldots,\psi_M,\phi_1,\ldots,\phi_L\}$.
It should be noticed that the construction (\ref{eq.expansion1})
ensures that the $M$ lowest eigenstates are contained in the space
spanned by the WFs.
Through the expansions (\ref{eq.expansion1}) and
(\ref{eq.expansion2}), $\mathcal{S}$ becomes a function of $U_{ij}$ and
$c_{ij}$, which must be minimized under the constraint that $U$
is unitary and the columns of $c$ are orthonormal. The minimization
can be performed by evaluating the derivatives of $\mathcal{S}$ with respect to 
$U_{ij}$ and $c_{ij}$ and then using any gradient based optimization
scheme to iteratively find the minimum. 
\end{subsection}

\begin{subsection}{Ethylene, $\text{C}_2\text{H}_4$}
  To illustrate the method we have constructed the WFs of an isolated
  ethylene molecule. We use a cubic supercell of length 14~\AA~and
  sample the BZ by the $\Gamma$ point. In Fig.~\ref{fig.c2h4WFs} we
  show iso-surfaces for two different sets of WFs corresponding to
  fully occupied WFs (upper row) generated with $E_0=E_F$ and $L=0$
  and partly occupied WFs generated with $E_0=E_F$ and $L=1$. In both
  cases we obtain four equivalent $\sigma$-orbitals located at the C-H
  bonds. In contrast, the WFs describing the C-C bond changes from two
  equivalent orbitals with a mixed $\sigma/\pi$ character, to a single
  $\sigma$ orbital centered between the C atoms and two atomic-like
  $p$-orbitals centered on each C. By inspection we find that the EDF
  selected in the minimization procedure in the latter case coincide
  with the anti-bonding molecular $\pi$-orbital of the C-C bond.
  Inclusion of this orbital is precisely what is needed to separate
  the $\sigma$- and $\pi$-systems on the molecule.
\begin{figure}[!ht]
\includegraphics[width=0.8\linewidth]{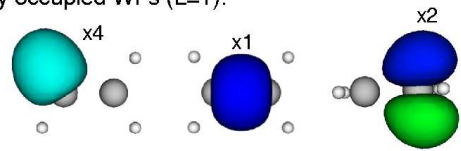}
\caption[cap.Pt6]{\label{fig.c2h4WFs} Occupied WFs (upper row) and
  partly occupied WFs (lower row) for ethylene, $\text{C}_2\text{H}_4$. The symbol $\times n$ 
  indicate the number of similar WFs obtained at equivalent sites on
  the molecule. Note, that by including the anti-bonding $\pi$ orbital
  in the WF expansion the $\sigma$- and $\pi$-systems become separated.}
\end{figure}
\end{subsection}

\begin{subsection}{Periodic systems}
We now turn to the construction of WFs for a periodic system. Suppose
that $\{w_n\}$
is a set of periodic functions defined in a cubic cell of
length $a$. It has been shown that in the limit of large $a$,
minimizing the spread (\ref{eq.spreadfct1}) is equivalent to maximizing
the functional~\cite{resta99}
\begin{equation}\label{eq.spreadfct2}
\Omega =\sum_{n}  (|X_{nn}|^2+|Y_{nn}|^2+|Z_{nn}|^2),
\end{equation}
where the matrix ${X}$ is defined as 
\begin{equation}\label{eq.xmatrix}
X_{nm}=\langle
w_n|e^{-i(2\pi/a)x}|w_m \rangle, 
\end{equation}
with similar definitions for $Y_{nm}$ and $Z_{nm}$. The generalization
to cells of arbitrary shape is straightforward and can be found
in Refs.~[\onlinecite{silvestrelli98,berghold00}]. Since an isolated system can always be studied in a large
periodic supercell within the $\Gamma$-point approximation, the
functional $\Omega$ can also be used instead of $\mathcal{S}$ to construct the
WFs of a finite system.

Since $\Omega$ corresponds to $\mathcal{S}$ only for periodic functions in large
supercells it seems to be useful only for systems that can be treated
within the $\Gamma$-point approximation. However, by using the folding
properties of the Brillouin zone (BZ) this problem can easily be
overcome. Consider a periodic system with a unit cell defined
by the basis vectors $\bold a_1,\bold a_2,\bold a_3$. The union
of the $\bold k$-subspaces associated with a uniform $N_1\times
N_2\times N_3$ grid in the 1.BZ which includes the $\Gamma$-point
coincides with the $\Gamma$-subspace of the repeated unit cell defined
by basis vectors $\bold A_i=N_i\bold a_i$.
Consequently, any function which can be written as a sum of functions
each characterized by a $\bold k$ from such a uniform grid is periodic
in the repeated cell and its spread can be defined by $\Omega$. In
this case the matrix elements (\ref{eq.xmatrix}) should be evaluated
over the repeated cell, however, as usual this can be reduced to
integrals over the small cell. Finally, we note that since the
number of $\bold k$-points determines the size of the repeated cell,
the original condition that the supercell should be sufficiently large
is turned into a condition of a sufficiently dense $\bold
k$-point sampling.

As for the non-periodic case, we start from a set of single-particle eigenstates,
$\{\psi_{n\bold k}\}$, obtained from a conventional electronic
structure calculation. For a periodic system, the eigenstates are Bloch
states labeled by a band index,
$n$, and a wave vector, $\bold k$. In accordance with the remarks
above the $\bold k$-points should belong to a uniform grid which includes 
the $\Gamma$-point.
We denote the total number of
bands used in the calculation by $N$, the desired number of WFs per
unit cell by $N_w$ and the number of eigenstates at a given $\bold
k$-point with energy below $E_0$ by 
$M_{\bold k}$. If $M_{\bold k}>N_w$, we set $M_{\bold k}=N_w$.
As in the isolated case, all eigenstates with energy
less than $E_0$ will be contained in the space spanned by the WFs.
We define the $n$th Wannier function belonging to cell
$i$ by
\begin{equation}\label{eq.periodicWF}
w_{i,n}=\frac{1}{\sqrt{N_k}}\sum_{\bold k} e^{-i\bold k \cdot \bold
  R_i}\tilde \psi_{n\bold k},
\end{equation}
where $N_k$ is the total number of $\bold k$-points and $\tilde
\psi_{n\bold k}$ is a generalized Bloch state to be defined below. 
We wish to find the set of generalized Bloch states that minimizes the
spread of the resulting Wannier functions. Since $w_{i,n}(\bold
r)=w_{0,n}(\bold r-\bold R_i)$ it is sufficient to consider only the
Wannier functions of the cell at the origin. 
For each of the generalized Bloch states we follow closely the idea
used in the isolated case. Thus we expand $\tilde
\psi_{n\bold k}$ in terms of the $M_{\bold k}$ lowest 
eigenstates and $L_{\bold k}=N_w-M_{\bold k}$ EDF,
$\{\phi_{l\bold k}\}$, from the remaining space of Bloch states:
\begin{equation}\label{eq.periodicexpansion1}
\tilde \psi_{n\bold k}=\sum_{m=1}^{M_{\bold k}}U(\bold k)_{mn}\psi_{m\bold
  k}+\sum_{l=1}^{L_{\bold k}}U(\bold k)_{M_{\bold k}+l,n}\phi_{l\bold k},
\end{equation}
where the unoccupied orbitals are written as
\begin{equation}\label{eq.periodicexpansion2}
\phi_{l\bold k}=\sum_{m=1}^{N-M_{\bold k}}c(\bold
k)_{ml}\psi_{M_{\bold k}+m,\bold k}.
\end{equation}
Since the states $\{\psi_{n\bold k}\}$ coincide with the $\Gamma$-point eigenstates of the
repeated unit cell we can use $\Omega$ to measure the spread of
$w_{0,n}$. The maximization of $\Omega$ with respect to $U(\bold
k)_{ij}$ and $c(\bold
k)_{ij}$ is performed in complete analogy with the isolated case.
For small systems ($\lesssim$ 10 atoms) the method is quite stable and usually leads to the same
global maximum independent of the initial guess used for the 
matrices $U(\bold
k)$ and $c(\bold k)$. In such cases the matrices can simply be initialized 
randomly. For larger systems, the method can get stuck in
local maxima and we have found it useful to start the maximization
from an initial guess of simple orbitals located either at the atoms
or the bond centers.
\end{subsection}

\begin{subsection}{Wannier function Hamiltonian}\label{sec.wfhamiltonian}
The transformation of eigenstates into localized WFs induces a
transformation of the Hamiltonian from its diagonal spectral
representation into a real-space or tight-binding like
representation. This real-space form of the Hamiltonian is in itself very
convenient as it provides direct access to physical parameters for
e.g. tight-binding models.

By combining Eqs.~(\ref{eq.periodicWF},\ref{eq.periodicexpansion1},\ref{eq.periodicexpansion2}) the
$n$th WF located in the unit cell at $\bold R_i$ can be compactly
written as
\begin{equation}\label{eq.simpelWF}
w_{i,n}=\frac{1}{\sqrt{N_k}}\sum_{\bold k,m}e^{-i\bold k \bold R_i}V(\bold k)_{mn}\psi_{m\bold k},
\end{equation}
where $V(\bold k)$ can be expressed in terms of $U(\bold k)$ and $c(\bold k)$.
The Hamiltonian matrix element
connecting the $n$th WF of cell $i$
with the $m$th WF of cell $j$ then becomes
\begin{eqnarray}\label{eq.WFhamiltonian}
H(\bold R_j-\bold R_i)_{nm}&\equiv &\langle
  w_{i,n}|H|w_{j,m}\rangle\nonumber \\&=&\sum_{\bold k,l}e^{-i(\bold R_j-
  \bold R_i)\cdot \bold k}V^*(\bold k)_{ln}V(\bold
  k)_{lm}\varepsilon_{l\bold k}\nonumber \\
\end{eqnarray}
where $\varepsilon_{l\bold k}=\langle \psi_{l\bold k}|H|\psi_{l\bold
  k}\rangle$ are the eigenvalues. Recall, that the WFs as defined here are
  periodic functions with a period given by the repeated cell. This
  periodicity is reflected in $H(\bold R_j-\bold R_i)_{nm}$. In order
  to describe fully localized functions the coupling matrix elements
  must therefore be truncated beyond a
  cut-off distance given approximately by $N_i/2$ unit cells in the direction $\bold a_i$.
This means in turn that the repeated cell should be large enough, or
  equivalently that the number of $\bold k$-points should be
  large enough, that the WFs have decayed sufficiently between the
  repeated images. 

To test the accuracy of the transformed WF Hamiltonian we can use it
to reproduce the original band structure. To this end we form the Bloch basis  
\begin{equation}\label{eq.TBbloch}
\chi_{n\bold k}=\frac{1}{\sqrt{N_R}}\sum_{\bold R_i}e^{i\bold R_i\cdot \bold k}w_{i,n},
\end{equation}
where $N_R$ is the number of unit cells included in the sum. 
The Hamiltonian matrix, $H(\bold k)$, in the Bloch basis for
a given $\bold k$ becomes
\begin{equation}\label{eq.approxeigenval}
\langle \chi_{n\bold k}|H|\chi_{m\bold k}\rangle=\sum_{\bold
  R_j}e^{i\bold k\cdot \bold R_j}H(\bold R_j)_{nm}.
\end{equation}   
The dimension of this matrix equals the number of WFs in one unit cell
and can be easily diagonalized yielding the approximate
eigenvalues $\tilde \varepsilon_{n\bold k}$. The quality of the
eigenvalues depends on the number of cells included in the
sum~(\ref{eq.approxeigenval}), i.e. on the range beyond which the
coupling is truncated. Since $H(\bold R_j)_{nm}$ becomes incorrect beyond
$N_i/2$ unit cells in direction $\bold a_i$ the quality of the
eigenvalues $\tilde \varepsilon_{n\bold k}$ ultimately depends on the
number of $\bold k$-points used in the construction of the WFs.  
\end{subsection}

\begin{subsection}{Trans-Polyacetylene}
As an example of a Wannier function analysis for a periodic system we
have constructed the WFs of trans-polyacetylene and generated the band
diagram using the corresponding WF Hamiltonian. We use a unit cell
containing two carbon and two hydrogen atoms and with a large amount of
vacuum separating the repeated wires by
more than $10$~\AA~in the directions perpendicular to the wire. We use a uniform $31\times
1\times 1 $ $\bold k$-point grid in the construction of the WFs. The
band diagram of the wire obtained directly from the electronic
structure calculation is shown in Fig.~\ref{fig.polyacetylene_band}
(full lines). For transport calculations we need a good description of
the electronic structure of the wire in an energy window around the
Fermi level, $\varepsilon_F=0$. Thus, in addition to the filled bands we
should describe at least the lower part of the $\pi^*$
conduction band correctly. Fig.~\ref{fig.polyacetylene_orbitals} shows
iso-surfaces of three different WFs obtained using the parameters
$N_w=6$ and $E_0=1.0$~eV. With this choice of parameters we obtain six WFs
per unit cell and ensure a correct description of the band structure
up to 1.0~eV. In total we obtain a $\sigma$-orbital at every C-H and C-C bond and an atomic-like
$p_z$-orbital centered on each C atom.
It is instructive to calculate the band structure of the wire using
these WFs as basis functions as described in section
\ref{sec.wfhamiltonian}. The result is shown by dots in
Fig.~\ref{fig.polyacetylene_band}. As expected the reconstructed bands
are in good agreement with the original bands below $E_0$. The high
density of bands starting around 3~eV above $\varepsilon_F$ marks the beginning of the
continuous spectrum. The fact that the reconstructed $\pi^*$ band does
not follow a particular band in this region indicates that the 
corresponding generalized Bloch states (\ref{eq.periodicexpansion1})
consist of non-trivial linear combinations of delocalized original
bands. This is sometimes referred to as band-disentanglement.

\begin{figure}[!ht]
\includegraphics[width=0.75\linewidth,angle=270]{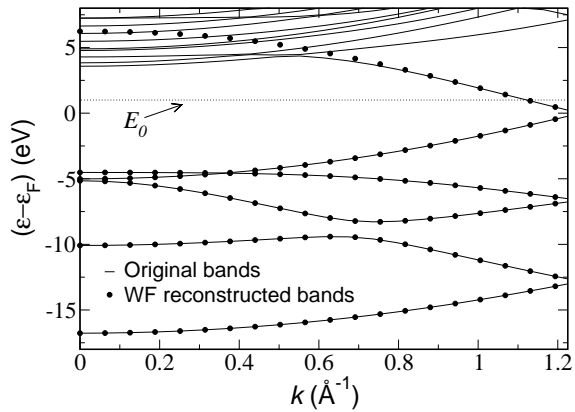}
\caption[cap.Pt6]{\label{fig.polyacetylene_band} Comparison between
  the band diagram obtained directly from the electronic structure
  calculation (full line) and the approximate band structure obtained
  using the WFs of Fig.~\ref{fig.polyacetylene_orbitals} as basis functions.}\end{figure}

\begin{figure}[!ht]
\includegraphics[width=0.42\linewidth]{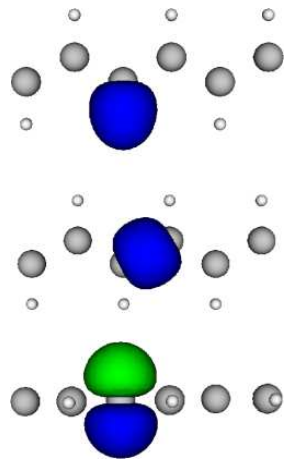}
\caption[cap.Pt6]{\label{fig.polyacetylene_orbitals} Partly occupied
  WFs for trans-polyacetylene. The unit cell contains two CH units.
  The WFs have been generated with the parameters $N_w=6$ and
  $E_0=1.0$~eV, where the Fermi level is set to 0~eV. There is one
  $\sigma$ orbital at every C-H and C-C bond and one atomic-like $p_z$
  orbital on each C.}
\end{figure}

\end{subsection}

\end{section}

\begin{section}{Representation of the Hamiltonian}\label{sec.hamiltonian}
In this section we link the WFs introduced in section~\ref{sec.WF} to 
the general transport formalism of section~\ref{sec.formalism}.
What is needed is a method to construct the
Hamiltonian~(\ref{eqn:scatter_hamiltonian}) for the coupled \mbox{$L$-$S$-$R$}
system in a localized WF basis set, and we will discuss a way of
achieving this. After a general discussion of some 
issues related to the use of supercells in connection with transport
calculations we discuss the construction of the three 
Hamiltonians $H_{\alpha},H_{S},H_{S\alpha}$. 

\begin{subsection}{Periodic supercells}
Within the periodic supercell approach, the system of interest is defined inside a finite
computational cell, the supercell, which is repeated in all directions to form a
super lattice. The principle is illustrated in   
Fig.~\ref{fig.bdtsupercell} for a 
single molecule suspended between infinitely extended surfaces. Clearly, we must
require that the
transverse dimensions of the supercell as well as the thickness of the
surface slabs are so large that the periodically repeated molecules do not interact.   
\begin{figure}[!ht]
\includegraphics[width=0.8\linewidth]{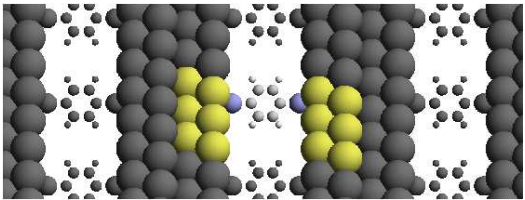}
\caption[cap.Pt6]{\label{fig.bdtsupercell} Illustration of a supercell describing a single molecule suspended
  between infinite surfaces.}
\end{figure}

Fig.~\ref{fig.bdtsupercell} shows an example of a 
scattering region as defined in the transport setup sketched in
Fig~\ref{fig.systemsetup}. If the surface slabs are
thick enough, the electron density and thus the
electron mean field potential at the supercell end-planes 
are close to their bulk values. It is therefore permissible to
extend the electron potential to the left and right of the supercell
by the bulk potential which leads to the system shown in   
Fig.~\ref{fig.bdtscatter}. We note, that the system remains periodic in
the directions perpendicular to the transport direction.
Within this approach, we are
therefore effectively considering transport through an infinite array of
contacts. Whether this provides a good description of transport
through a single, isolated contact then depends on the degree of interference
between the repeated contacts. As the transverse dimensions of the supercell are increased
this interference is expected to decrease and
the result should approach that of a single contact.

Since the scattering region of the system shown in
Fig.~\ref{fig.bdtscatter} consists of an infinite array of
molecules, all matrices entering the transmission
formula~(\ref{eq.transmission}) have infinite dimensions.
However, the periodicity of the system implies that the wave vectors, $\bold
k_{\perp}$, belonging to the 1.BZ. of the transverse plane are good quantum
numbers. Consequently, all the matrices entering
Eq.~(\ref{eq.transmission}) are diagonal with respect to $\bold
k_{\perp}$ and the total transmission can thus be decomposed into a sum
over $\bold k_{\perp}$-dependent transmissions given by
\begin{equation}
T(\bold k_{\perp};\varepsilon)=\text{Tr}\left[
\boldsymbol{G}_{S}^{r}(\bold k_{\perp};\varepsilon)
\boldsymbol{\Gamma}_{L}(\bold k_{\perp};\varepsilon)
\boldsymbol{G}^{a}_{S}(\bold k_{\perp};\varepsilon)
\boldsymbol{\Gamma}_{R}(\bold k_{\perp};\varepsilon)
                  \right].
\end{equation}
The transmission per
supercell is evaluated as $T(\varepsilon)=\sum_{\bold
  k_{\perp}}W_{\bold k_{\perp}}T(\bold k_{\perp};\varepsilon)$,
where $W_{\bold k_{\perp}}$ are appropriate weight factors adding up
to 1.
\begin{figure}[!ht]
\includegraphics[width=0.75\linewidth]{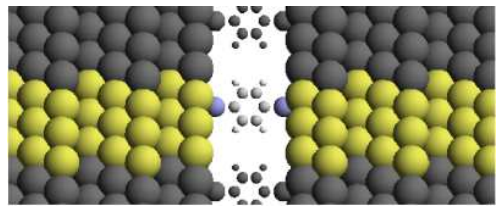}
\caption[cap.Pt6]{\label{fig.bdtscatter} The system obtained by
  extending the electron potential to the left and right of the supercell shown in
  Fig.~\ref{fig.bdtsupercell} by the bulk
  potential. Note, that the
  system remains periodic in the directions perpendicular to the
  transport direction.}
\end{figure}
\end{subsection}

\begin{subsection}{Some notation}
  For use in the following we introduce some notation. We take the
  direction of transport to coincide with the $z$-axis. For any system
  of the form illustrated in Figs.~\ref{fig.bdtsupercell},\ref{fig.bdtscatter} we denote the basis vectors of the super
  lattice in the transverse plane by $\bold A_1$ and $\bold A_2$. We
  assume that this plane is perpendicular to the transport direction,
  i.e. $\bold A_1\cdot \bold e_z=\bold A_2\cdot \bold e_z=0$, and
  refer to the unit cell spanned by $\bold A_1$ and $\bold
  A_2$ as the transverse cell. A general lattice vector within this
  plane is denoted by $\bold R_{\perp}$, and can thus be written $\bold R_{\perp}=n\bold
  A_1+m\bold A_2$, where $n,m$ are integers. Finally, we denote the
  wave-vectors in the 1.BZ of the transverse cell by $\bold
  k_{\perp}$.
\end{subsection}

\begin{subsection}{Lead Hamiltonian ($\boldsymbol{H}_{\alpha}$)}
In the following we describe how to construct 
the Hamiltonian matrix, $\boldsymbol{H}_{\alpha}$, for
lead $\alpha$. From
Eq.~(\ref{eqn:leadgfeq}) it is clear that we need 
the matrices $\boldsymbol{h}_0$ and $\boldsymbol{h}_1$ describing a single
principal layer and the coupling between two neighboring layers, respectively.

Suppose that a single principal layer of lead $\alpha$ is described
by a supercell with basis vectors $\bold A_i=N_i\bold a_i$, $i=1,2,3$.
If $N_i>1$ for some $i$, the principal layer can be build from a smaller unit cell with
basis vectors $\bold a_1,\bold a_2,\bold a_3$. In the example
illustrated in Fig.~\ref{fig.bdtsupercell} a possible principal layer super
cell consists of one ABC period of $3\times 3$ sections of the (111) atomic planes of
the bulk fcc crystal. This cell can in turn be made up of smaller cells
containing only $1\times 1$ sections in the transverse plane. Let
$\{w_{i,n}\}$, $n=1,\ldots,N_w$ denote the set of WFs obtained for the small
cell at $\bold R_i$. For a given $\bold k_{\perp}$ we form the
Bloch functions
\begin{equation}\label{eq.blochbasis}
\chi_{i,n}(\bold k_{\perp})=\frac{1}{\sqrt{N_{R_{\perp}}}}\sum_{\bold
  R_{\perp}}e^{i\bold k_{\perp}\cdot \bold R_{\perp}}w_{i,n}(\bold
  r-\bold R_{\perp}),
\end{equation}
 where $N_{R_{\perp}}$ is the number of transverse cells included in the
 sum and $i$ denotes a small unit cell. Note, that even though the
 Bloch functions are extended in the transverse directions they
 preserve the finite range of the WFs in the transport direction. The corresponding Hamiltonian
 matrix is given by 
\begin{eqnarray}
H(\bold k_{\perp})_{in,jm}&\equiv&\langle \chi_{i,n}(\bold
k_{\perp})|H|\chi_{j,m}(\bold k_{\perp})\rangle \nonumber\\
&=&\sum_{\bold
  R_{\perp}} e^{i\bold k_{\perp}\cdot \bold R_{\perp}}H(\bold
  R_j-\bold R_i+\bold R_{\perp})_{nm},
\end{eqnarray}
where the matrix elements $H(\bold
  R_j-\bold R_i+\bold R_{\perp})_{nm}$ refer to the WF Hamiltonian of
  Eq.~(\ref{eq.WFhamiltonian}). To obtain the matrix $\boldsymbol{h}_0$ we let $i,j$
  run over all small unit cells within one principal layer supercell
  and $n,m$ run over all WFs. To obtain $\boldsymbol{h}_1$ we do the same except that
  $j$ now runs over all small unit cells within the supercell of the neighboring
  principal layer. The accuracy of the Hamiltonian $\boldsymbol{H}(\bold k_{\perp})$ depends on the
number of transverse cells included in the sum. As explained in
  sec.~\ref{sec.wfhamiltonian} this number is limited by the size
  of the period of the WFs which in turn is determined by the number of $\bold
  k$-points used to construct them. For a transverse cell
  containing $3\times 3$ atoms we have found it sufficient to sum over
nearest neighbor cells.
\end{subsection}

\begin{subsection}{Hamiltonian for scattering region ($\boldsymbol{H}_{S}$) and coupling ($\boldsymbol{H}_{S\alpha}$) }
  The technique for constructing a matrix representation of the lead
  Hamiltonian described in the previous section can also be applied to
  obtain the Hamiltonian for the scattering region,
  $\boldsymbol{H}_{S}$. However, the coupling between $S$ and the
  leads, $\boldsymbol{H}_{S\alpha}$, involves matrix elements
  connecting WFs in the lead with WFs in $S$, and since these are in
  general different the method described above does not apply to the
  coupling matrix. Instead, we use the electronic
  structure code to directly evaluate these matrix elements.

We define a composite supercell, $\widetilde S$, by attaching one lead principal layer
cell on each side of the supercell containing $S$. The electron
density and the atomic configuration in the
new supercell are obtained by "gluing" together the densities and
atomic configurations of the three constituent parts. Since the
electron density defines the effective potential, this determines
the Hamiltonian $H_{\widetilde S}$.

Let us denote by $\mathbb B_1$ the combined set of WFs for $S$ and 
the two principal layers closest to $S$. 
For each WF in $\mathbb B_1$ we form the Bloch 
function~(\ref{eq.blochbasis}) on a real space grid in the composite
supercell $\widetilde S$. The resulting set of Bloch functions is denoted $\mathbb
B_2$. Note, that the Bloch functions in the lead parts of $\widetilde
S$ are identical to those used to construct the lead Hamiltonian.
Since the WFs are originally defined in separate simulation
cells, the introduction of these WFs into the composite cell $\widetilde S$
involve some manipulations which are easiest to
implement in real space. For illustration, consider a WF of $S$
constructed using only a single $\bold k$-point in the transport
direction. Since the period of the WF is given by the length of $S$,
it cannot be directly introduced in $\widetilde S$. Instead, we first
translate it to the center of $S$ (using periodic boundary
conditions within $S$), then copy it into the $S$ part
of $\widetilde S$ and extend it into the lead parts of $\widetilde
S$ by zero-"tails". Finally, the WF is translated back to its original position (using
periodic boundary conditions within $\widetilde S$). 

Given the set of Bloch functions, $\mathbb B_2$, defined in the
composite supercell as well as the Hamiltonian, $H_{\widetilde S}$, we use the
electronic structure code to evaluate the Hamiltonian matrix
elements. This results in a matrix of the form
\begin{equation}
\label{eqn:scatter_hamiltonian2}
\boldsymbol{H}_{\widetilde S}=\left(
                \begin{array}{ccc}
                \tilde{\boldsymbol{h}}_0 &
                \boldsymbol{H}_{SL}^{\dagger} & \tilde{\boldsymbol{h}}_1^{\dagger} \\
                \boldsymbol{H}_{SL} & \boldsymbol{H}_{S} & \boldsymbol{H}_{SR} \\
                \tilde{\boldsymbol{h}}_1 &
                \boldsymbol{H}_{SR}^{\dagger} & \tilde{\boldsymbol{h}}_0 \\
                \end{array}
               \right).
\end{equation}
The matrices $\tilde{\boldsymbol{h}}_1$ and $\tilde{\boldsymbol{h}}_1^{\dagger}$ are due to the periodic boundary
conditions. Except for a constant, the matrix
$\tilde{\boldsymbol{h}}_0$ should equal $\boldsymbol{h}_0$ and a
comparison of the two can serve as a consistency check. Finally, 
$\boldsymbol{H}_{\widetilde S}$ must be shifted by a constant to
account for the unspecified energy-zero and thereby ensure a
smooth matching at the interface between lead and scattering region.
This can be done by adding to $\boldsymbol{H}_{\widetilde S}$ the
matrix $c\boldsymbol{S}_{\widetilde S}$, where $c=[\boldsymbol{h}_0]_{00}-[\tilde{\boldsymbol{h}}_0]_{00}$.
\end{subsection}
\end{section}

\begin{section}{Results and Analysis}\label{sec.results}
In this section we apply the transport scheme to two molecular
contacts: a hydrogen molecule between atomic Pt chains and
benzene-dithiol between Au(111) surfaces. We introduce a useful
technique for relating the various features of the transmission
function to specific molecular orbitals. This enables us to identify
the anti-bonding $\text{H}_2$ orbital as the current carrying state,
and to explain the transmission through the benzene-dithiol in terms
of only two molecular orbitals. 

\begin{subsection}{Molecular hydrogen bridge in a Pt chain}
  As a first example we study the transmission through a molecular
  hydrogen bridge between a pair of semi-infinite, monatomic Pt
  chains, see Fig.~\ref{fig.ptwire}(a). In a recent experiment the
  conductance of a molecular hydrogen bridge suspended between bulk Pt
  electrodes was measured using mechanically controlled break
  junctions~\cite{smit02}.  The system was found to have a conductance
  close to the conductance quantum, $G_0=2e^2/h$. While this has been
  confirmed by at least three different
  calculations~\cite{smit02,cuevas_heurich03,thygesen_h2}, there is at
  least one calculation predicting a conductance of only
  $0.2G_0$~\cite{garcia_palacios04}. For simplicity we use monatomic
  chain leads where the transverse $\bold k$-point sampling can be
  safely omitted.

We represent the system in a supercell with transverse dimensions
$10$\AA$\times 10$\AA. The relevant bond lengths are:
$d_{\text{Pt-Pt}}=2.41$~\AA, $d_{\text{Pt-H}}=1.67$~\AA~and
$d_{\text{H-H}}=0.88$~\AA. We include 4 Pt atoms in a principal layer
and define the scattering region as 4Pt$-\text{H}_2-$4Pt. 

\begin{figure}[!ht]
\includegraphics[width=0.85\linewidth]{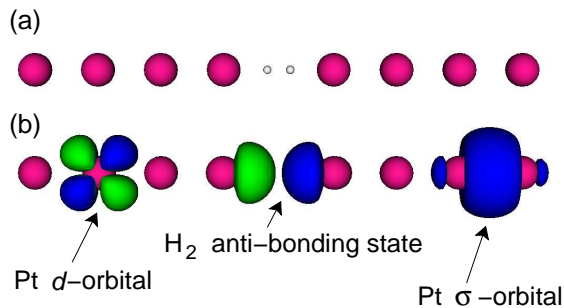}
\caption[cap.Pt6]{\label{fig.ptwire} (a) Molecular hydrogen bridge
  connecting two semi-infinite, monatomic Pt chains. The atoms shown
  constitute the scattering region in the conductance calculation. (b)
  Contour plots of three representative basis
  functions. After diagonalizing the molecular subspace, the basis set
  consists of the initial WFs in the lead region and renormalized $\text{H}_2$ bonding and
  anti-bonding states in the molecular region.}
\end{figure}

In Fig.~\ref{fig.ptwiretrans} we show the calculated 
transmission (thick line) as a function of energy. The transmission vanishes below
$-6.5$~eV which marks the bottom of the $s$-band of the
infinite Pt chain. The predicted conductance, $G=T(\varepsilon_F)$, is
found to be close to the experimental value of $1G_0$.
Ideally, we would like to understand the shape of the
transmission curve in terms of the properties of the uncoupled
systems, i.e. the isolated $\text{H}_2$ molecule and the free leads.
The interaction of a molecule with the leads can be seen as a
two-step process: first the molecular levels are shifted
(renormalized) due to the change in the potential from the
leads. Next, the renormalized levels hybridize with the states in the
leads as a consequence of the overlap between the wave functions.
In the following we describe how the first step can be
implemented and studied in practice. We will return to the second
step in Sec.~\ref{sec.bdt}.

\begin{figure}[!ht]
\includegraphics[width=0.8\linewidth,angle=270]{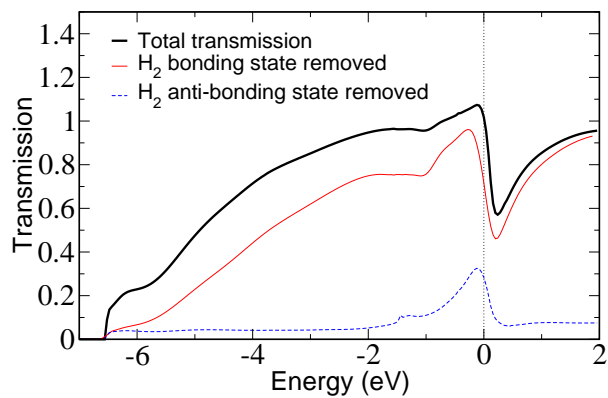}
\caption[cap.Pt6]{\label{fig.ptwiretrans} Transmission through the
  molecular hydrogen bridge shown in Fig.~\ref{fig.ptwire}(a). The
  thick line represents the total transmission while the dashed and
  thin lines refer to the situations where the $\text{H}_2$ bonding, respectively
  anti-bonding states have been removed.}
\end{figure}

\begin{subsubsection}{Diagonalization of the molecular subspace}\label{sec.diagonal}
In order to obtain the renormalized energy levels and orbitals of a
molecule in contact with leads, we first construct the WFs for
the combined system, i.e. the scattering region, and the corresponding
Hamiltonian matrix, $\boldsymbol{H}_S$. Next, we identify the set of
WFs located on the molecule. This involves some arbitrariness
since it is not always clear where to put the separation between
molecular WFs and lead WFs. Usually, the distinction is clearer when
the bonding between molecule and lead is weaker.
We refer to the space spanned by
the molecular WFs as the molecular subspace.
The renormalized molecular levels and orbitals are obtained by
diagonalizing $\boldsymbol{H}_S$ within the molecular subspace. 
Since molecular orbitals to a large extent are determined by the symmetry of
the molecule, it is usually straightforward to relate the 
renormalized orbitals to the orbitals of the free molecule.
In contrast, the renormalized energies
can be drastically shifted compared to the energies of the free molecule. 
\end{subsubsection}

In the case of $\text{H}_2$ the molecular diagonalization produces a
bonding, $|b\rangle$, and anti-bonding, $|a\rangle$, state from the initial $1s$-like WFs centered
on each H. The renormalized energies are 
$\varepsilon_b=\langle b|H|b\rangle=-7.4$~eV~and
$\varepsilon_a=\langle a|H|a\rangle=0.4$~eV, relative to the Fermi
level of the infinite chains. For comparison the corresponding energy levels of the
free molecule before coupling are $-4.2$~eV and $6.6$~eV,
respectively. The anti-bonding state is thus shifted down close to the
Fermi level. This agrees with the conventional understanding of
hydrogen dissociation on transition metal surfaces which has 
been established on the basis of DFT calculations~\cite{hammer95}. In
fact the filling of the $\text{H}_2$ anti-bonding state weakens the
hydrogen bond and eventually leads to dissociation. The effect is less
dramatic for hydrogen in the contact, however, the fractional filling
of the anti-bonding state does enlarge the H-H bond length as compared
to the free molecule. The reduction of the HOMO-LUMO gap from 10.8~eV
in the free molecule to 7.8~eV in the contacted molecule is partly due
to this structural effect ($\approx
1$~eV). The remaining reduction is due to the lead-induced change
of the potential around the molecule. In Fig.~\ref{fig.ptwire}(b) we show contour plots of $|a\rangle$
together with Pt $d$- and $\sigma$-like WFs.

Having obtained the renormalized molecular orbitals, it is natural to
ask if there is significant interference between the two levels or if
the conduction properties of the system are determined by just one of
the two orbitals. A very direct way to answer this question, is to
calculate the transmission with either the bonding or anti-bonding
state removed. In practice this is done by setting all Hamiltonian
matrix elements involving either $|b\rangle$ or $|a\rangle$ to zero.
The resulting transmission curves are shown in
Fig.~\ref{fig.ptwiretrans}. It is clear that the transmission is
mainly due to the anti-bonding state and that interference effects are
not very pronounced.  The small peak in the transmission through the
bonding state around the Fermi level is due to a peak at the same
position in the DOS of the group orbital of $|b\rangle$ (see
Sec.~\ref{sec.newns}).  Although the use of monatomic chains as leads
is an over-simplification, the main conclusions regarding the value of
the conductance, the position of the molecular levels and the
mechanism of conduction are unchanged when realistic bulk electrodes
are used~\cite{thygesen_h2}.
\end{subsection}

\begin{subsection}{Benzene-dithiol between Au(111) surfaces}\label{sec.bdt}
In this section we study the transmission through a 
benzene-dithiol molecule (BDT) suspended between Au(111)
surfaces. The transport properties of this system have been
investigated experimentally by Reed et al.~\cite{reed97} using the
mechanically controllable break junction
technique, and several theoretical studies  
have been reported subsequently~\cite{emberly01,diventra00,derosa01,stokbro03,xue_quantumchem}. 
All of these
studies are, like the present, based on a combination of the Landauer
B{\"u}ttiker formalism where a DFT description of the electronic
structure enters at different levels. In general, the calculations 
for a BDT molecule covalently bonded to the electrodes finds a
conductance which is 2-3 orders of
magnitude higher than the experimental value. Different reasons for
this discrepancy have been proposed including overlapping BDT
molecules~\cite{emberly01} and attachment of the BDT on top of a gold atom~\cite{diventra00}.
Here we focus on a symmetrically, covalently bonded molecule. We  
find a conductance comparable to those obtained in previous
theoretical studies, however, we illustrate in addition how
a proper construction of renormalized molecular orbitals leads to  
a very simple picture of the transport mechanism in terms of
only two molecular states. The relatively large number
of conductance calculations reported for the BDT system 
makes it an ideal test system 
for comparing different computation methods.

We use a supercell containing $3\times 3$ Au atoms in the directions
perpendicular to the transport. The lead principal layer 
contains three atomic planes in accordance with the ABC-stacking. The scattering region contains 3 Au layers on each side of
the DTB molecule which is enough to achieve a smooth matching with the
bulk potential. To obtain a contact geometry which more closely
resembles the open structures likely to occur in the
experiments we insert a 3-atom pyramid between the surface and the molecule. 
The central part of the scattering region is shown in
Fig.~\ref{fig.bdt_orbitals}(a). We use the bond lengths
$d_{\text{Au-S}}=2.40$~\AA, $d_{\text{S-C}}=2.75$~\AA,
$d_{\text{C-C}}=1.40$~\AA~and $d_{\text{C-H}}=1.08$~\AA~in accordance
with those used in Ref.~\onlinecite{stokbro03}. All gold atoms
are fixed in an fcc lattice with lattice parameter $a=4.18$~\AA.

\begin{figure}[!ht]
\includegraphics[width=0.77\linewidth]{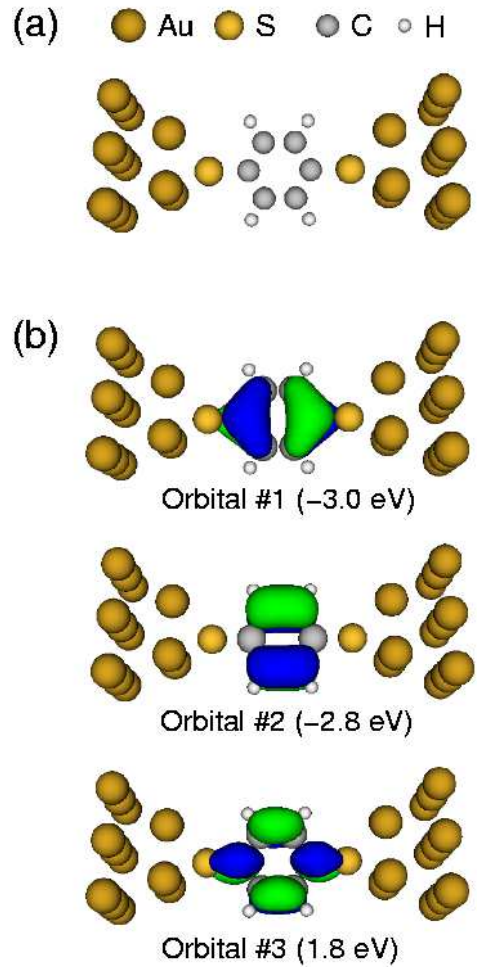}
\caption[cap.Pt6]{\label{fig.bdt_orbitals} (a) Benzene-dithiol (BDT)
  between Au(111) surfaces. (b) The three renormalized molecular orbitals
  closest to the Fermi level of the electrodes. The renormalized
  energy is indicated in parenthesis. Orbital 2 (HOMO) has negligible
  weight on the S atoms and has no influence on the conduction
  properties. In contrast orbitals 1 and 3 are strongly coupled to the lead
  states and in fact they completely determine the transmission function.}
\end{figure}

In Fig.~\ref{fig.bdt_trans} we show the calculated transmission
obtained using 1, 8 and 18 irreducible $\bold k$-points to sample the
transverse BZ. The $\Gamma$-point transmission fluctuates very much
compared to the converged result and is clearly not a good
approximation. The rapid fluctuations are characteristic of
$\Gamma$-point transmissions. This is because the lead in this case is
simply a one-dimensional pipe with a $3\times 3$-atom cross section.
The electronic structure of such relatively thin $1d$ structures is
very different from the electronic structure of bulk systems, in
particular the DOS of the former exhibits van Hove singularities at
the $1d$ band edges~\cite{thygesen_kpoints}. We mention that the band
dispersion between the free BDT molecules is less than 0.1~eV for all
relevant orbitals. This shows that the $\bold k$-point dependence of
the transmission function is not due to direct coupling between the
repeated molecules, but is rather a substrate induced effect.  The
calculated transmission shows two broad peaks at around $-3.0$~eV and
$2.5$~eV, and two more narrow peaks around $-1.5$~eV and $-1.0$~eV
(all energies are with respect to the Fermi level of the leads). The
occurrence of two broad peaks, one below and one above the Fermi
level, is in qualitative agreement with previous
results~\cite{emberly01,stokbro03,xue_quantumchem}. The position of
these peaks as well as the existence of the two narrow peaks just
below the Fermi level, are not in agreement with the earlier
calculations which also deviate somewhat from each other.  These
discrepancies might be due to differences in the adopted geometries or
technical issues like adequate $\bold k$-point sampling. In
particular the result in Ref.~\onlinecite{emberly01} has been averaged
over different atomic geometries which might smear out features
related to details in the lead electronic structure. As we shall see,
the narrow peaks arise exactly due to such details.  Finally we note,
that the predicted conductance is around $0.1G_0$. This is in very
good agreement with the results obtained in
Refs.~\onlinecite{emberly01,xue_quantumchem}, while it is a factor of
4 smaller than the result of Ref.~\onlinecite{stokbro03} and a factor
of 3 larger than the result of Ref.~\onlinecite{diventra00}.

\begin{figure}[!ht]
\includegraphics[width=0.8\linewidth,angle=270]{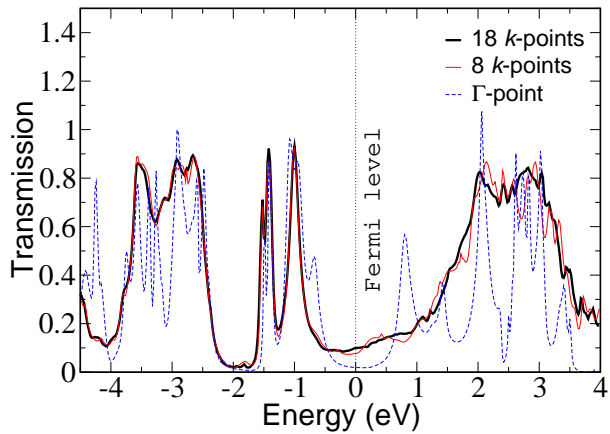}
\caption[cap.Pt6]{\label{fig.bdt_trans} Calculated transmission
  through the bdt-system shown in Fig.~\ref{fig.bdt_orbitals} for
  three different $\bold k$-point samplings of the transverse BZ.}
\end{figure}

To investigate which orbitals of the BDT molecule are responsible for
the different features of the transmission function we have performed a
diagonalization of the molecular subspace as described in
Sec.~\ref{sec.diagonal}. We define the molecular subspace as those WFs
whose centers lie closer to the $\text{C}_6\text{H}_4$ molecule than
to any other atom. The reason why we do not include the WFs at the S
atoms in the molecular subspace is that S is three-fold coordinated to
the gold and the "weak" link in the contact is more naturally defined
between S and C. Diagonalizing the Hamiltonian within the molecular
subspace leads to 16 renormalized orbitals distributed in the energy
range -17~eV to 5~eV. In Fig.~\ref{fig.bdt_orbitals}(b) we show
three renormalized molecular orbitals with energy close to the Fermi
level. These orbitals all belong to the $\pi$-system of the molecule,
i.e. they change sign under a reflection in the molecular plane.

Orbital 2 has negligible weight in the vicinity of the S atom
and is expected to couple only weakly to the lead states. In contrast
orbitals 1 and 3 have significant weight at the S-C bond and
should therefore couple strongly with the leads and contribute
correspondingly to the transmission.  To obtain more quantitative
information we have calculated the transmission function with all
molecular orbitals except orbital 1, respectively 3, removed from the
basis set, see Fig.~\ref{fig.bdt_trans1+3}. The result very clearly
demonstrates that the broad peak at $-3$~eV is exclusively due to
orbital 1, while the peak at $2.5$~eV is exclusively due to orbital 3.
In particular interference between the different molecular orbitals
does not occur around these energies. We can also conclude that the
narrow peak at $-1.5$~eV is mostly due to orbital 3 while the narrow
peak at $-1.0$~eV is mostly due to orbital 1, although the separation
is not so clear in this case and interference effects do play a role.

\begin{figure}[!ht]
\includegraphics[width=0.8\linewidth,angle=270]{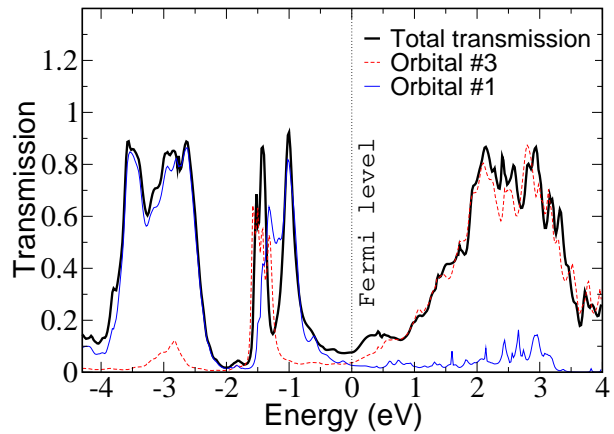}
\caption[cap.Pt6]{\label{fig.bdt_trans1+3} Calculated transmission
  through the bdt-system when all molecular orbitals except from orbital 1 (thin) and orbital 3
  (dashed) have been removed. The total transmission (thick) is shown
  as a reference. All calculations have been performed using 8
  irreducible $\bold k$-points. Notice, that the main transmission
  peak at $-3$~eV ($2$~eV) can be directly related to orbital 1 (3).}
\end{figure}

The above analysis shows that the transport properties of the BDT are
completely determined by orbitals 1 and 3. To take the analysis even
further we have constructed the corresponding group orbitals (see
Sec.~\ref{sec.newns}), both of which are quite similar to an
S-centered $p$-orbital perpendicular to the molecular plane. This is
easily anticipated from the $\pi$ character of the molecular orbitals. According
to Sec.~\ref{sec.newns} the DOS of the group orbital calculated
without coupling to the molecule, together with the level position and
the coupling matrix element, determines the projected DOS for the
corresponding molecular orbital.  The group orbital DOS is shown in the upper panel of
Fig.~\ref{fig.bdt_group} (dashed line). We can model the DOS by a
semi-elliptical band on top of a flat background (solid line), which
allows us to compute the resulting DOS of an orbital with energy
$\varepsilon_a$ and coupling $V$. The coupling can be directly
extracted from the DFT Hamiltonian giving $V=1.3$~eV. Using the
on-site energies of the renormalized orbitals 3 and 1, we obtain the
DOS shown in the middle and bottom panel, respectively. In addition to
the broad resonances formed close to the bare on-site energies, a
narrow bonding, respectively anti-bonding resonance, involving the
molecular orbital and the lead states, is formed at the edges of the
semi-elliptical band. On the basis of these observations and
the conductance formula Eq.~(\ref{eq.newns}) we can finally conclude that
the transmission peaks at $2.5$~eV and $-1.5$~eV are due to orbital 3,
while the peaks at $-3.0$~eV and $-1.0$~eV are due to orbital 1.

It is worth noting that the simplicity of the picture described above
in terms of only two molecular orbitals relies on our 
definition of the molecular subspace. In
Refs.~\onlinecite{stokbro03,derosa01} the transmission was analyzed in terms
of the molecular orbitals of
the whole BDT molecule~\cite{stokbro03}, i.e. including the S atoms,
and the BDT molecule plus selected Au atoms~\cite{derosa01}. This leads to a
more complicated description of the transmission peaks involving
several different molecular orbitals. 
Thus the precise division of the Hilbert space
into molecule- and lead-subspaces can lead to more or less elegant descriptions
of transport properties. 

\begin{figure}[!ht]
\includegraphics[width=0.8\linewidth,angle=270]{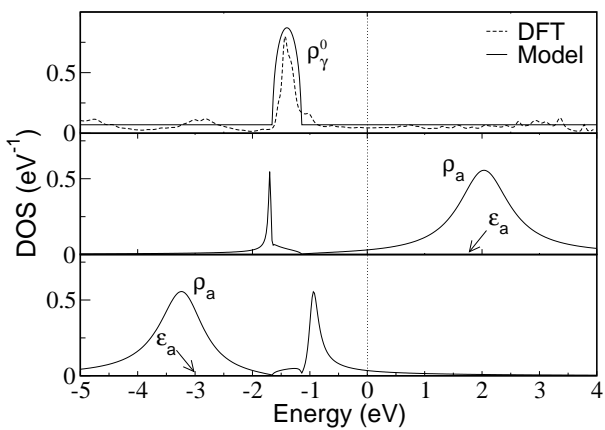}
\caption[cap.Pt6]{\label{fig.bdt_group} The upper panel shows the DOS, $\rho_{\gamma}^0$,
  of the (common) group orbital of molecular orbitals 1 and
  3. $\rho_{\gamma}^0$ has been calculated with all coupling matrix
  elements involving the molecular orbitals set to 0, i.e. it refers
  to the uncoupled leads. The dashed line is the result obtained from
  the calculation, while the full line is a model fit. The two lower
  panels show the resulting DOS of a level }
\end{figure}

\end{subsection}

\end{section}

\begin{section}{Conclusions}\label{sec.conclusions}
We have presented a numerical method for calculating
electron transport in atomic-scale contacts. The method combines a
formally exact Green's function formalism with a mean-field
description of the electronic structure based the Kohn-Sham scheme of
density functional theory. 
By transforming the delocalized 
eigenstates obtained from an accurate plane-wave calculation, into
maximally localized Wannier functions we obtain a highly efficient
minimal basis set for evaluating the relevant Green's functions.
The construction scheme used to obtain the WFs has been introduced
for both isolated and periodic systems and examples have been given to
illustrate the important feature of bonding-antibonding
closure. Finally, we have applied the transport scheme to a molecular
hydrogen contact in a monatomic Pt wire and a benzene-dithiol molecule
between Au(111) surfaces. A useful analysis technique for identifying
which molecular orbitals contribute to which features of the
transmission function has been
introduced and applied to each of the systems. In this way we showed
that the transport through the hydrogen molecule is determined by the
anti-bonding state, and we identified two molecular orbitals which
completely accounts for the transport through the benzene-dithiol.
 
\end{section}

\begin{section}{Acknowledgments}\label{sec.acknowledgments}
We thank Lars B. Hansen for contributions to the Wannier function scheme.
We acknowledge support
from the Danish Center for Scientific Computing through Grant No.
HDW-1101-05.
\end{section}

\newpage


\bibliographystyle{apsrev}

\end{document}